\documentclass[aps,pra,superscriptaddress,twocolumn]{revtex4}%
\usepackage{graphics}
\usepackage{amsmath}
\usepackage{graphicx}
\usepackage{amsfonts}
\usepackage{amssymb}%
\begin{document}
\title{MDI-QKD: Continuous- versus discrete-variables at metropolitan distances}

\begin{abstract}
In a comment, Xu, Curty, Qi, Qian, and Lo claimed that discrete-variable (DV)
measurement device independent (MDI) quantum key distribution (QKD) would
compete with its continuous-variable (CV) counterpart at metropolitan
distances. Actually, Xu et al.'s analysis supports exactly the opposite by
showing that the experimental rate of our CV protocol (achieved with practical
room-temperature devices) remains one order of magnitude higher than their
purely-numerical and over-optimistic extrapolation for qubits, based on
nearly-ideal parameters and cryogenic detectors (unsuitable solutions for a
realistic metropolitan network, which is expected to run on cheap
room-temperature devices, potentially even mobile). The experimental rate of
our protocol (expressed as bits per relay use) is confirmed to be two-three
orders of magnitude higher than the rate of any realistic simulation of
practical DV-MDI-QKD over short-medium distances. Of course this does not mean
that DV-MDI-QKD\ networks should not be investigated or built, but increasing
their rate is a non-trivial practical problem clearly beyond the analysis of
Xu et al. Finally, in order to clarify the facts, we also refute a series of
incorrect arguments against CV-MDI-QKD and, more generally, CV-QKD, which were
made by Xu et al.\ with the goal of supporting their thesis.

\end{abstract}
\author{Stefano Pirandola}
\affiliation{Computer Science and York Centre for Quantum Technologies, University of York,
York YO10 5GH, United Kingdom}
\author{Carlo Ottaviani}
\affiliation{Computer Science and York Centre for Quantum Technologies, University of York,
York YO10 5GH, United Kingdom}
\author{Gaetana Spedalieri}
\affiliation{Computer Science and York Centre for Quantum Technologies, University of York,
York YO10 5GH, United Kingdom}
\author{Christian Weedbrook}
\affiliation{Department of Physics, University of Toronto, Toronto M5S 3G4, Canada}
\author{Samuel L. Braunstein}
\affiliation{Computer Science and York Centre for Quantum Technologies, University of York,
York YO10 5GH, United Kingdom}
\author{Seth Lloyd}
\affiliation{Department of Mechanical Engineering and Research Laboratory of Electronics,
Massachusetts Institute of Technology, Cambridge MA 02139, USA}
\author{Tobias Gehring}
\affiliation{Department of Physics, Technical University of Denmark, Fysikvej, 2800 Kongens
Lyngby, Denmark}
\author{Christian S. Jacobsen}
\affiliation{Department of Physics, Technical University of Denmark, Fysikvej, 2800 Kongens
Lyngby, Denmark}
\author{Ulrik L. Andersen}
\affiliation{Department of Physics, Technical University of Denmark, Fysikvej, 2800 Kongens
Lyngby, Denmark}
\maketitle


In a recent comment~\cite{Fehiu}, Xu et al.\ claimed that discrete variable
(DV) measurement device independent (MDI) quantum key distribution
(QKD)~\cite{SidePRL,Lo,Note} was unfairly compared to a novel high-rate
continuous variable (CV) protocol~\cite{CVMDIQKD}. Here we show that this
claim is false and we fully clarify this DV-CV comparison. However, before
going into the details of this comparison, we need to rectify a series of
incorrect and misleading statements made by these authors against CV-MDI-QKD
and, more generally CV-QKD, with the aim of supporting their thesis.

\section*{Features of CV-MDI-QKD}

First of all, contrary to the claims of Xu et al.~\cite{Fehiu}, the CV
experiment of~\cite{CVMDIQKD} is performed with cheap room-temperature
components (optical modulators and homodyne detectors) in a regime of
parameters which are easily achievable in practice. Modulations of
$\varphi\simeq60$ shot noise units are relatively low with respect to what is
achievable ($100$ and more). Reconciliation efficiencies of $\xi\simeq97\%$
are currently state-of-the-art in CV-QKD
experiments~\cite{Jouguet,Elkouss,RMP}. An experimental excess noise
$\varepsilon\simeq0.01$ is not low but typical, fully comparable with the
values reported in the fibre-optic experiment of~\cite{Grangier2}, where the
experimental excess noise was estimated to be $\varepsilon\simeq0.001$ at
$10^{8\text{ }}$data points and $\varepsilon\simeq0.008$ at $10^{6\text{ }}%
$data points, for Bob's detection at $53$~km. At lower distances $\lesssim
25$~km (as is in our case), the excess noise is expected to be smaller, which
means that our experimental value $\varepsilon\simeq0.01$ can even be
considered relatively high. The robustness of CV-MDI-QKD against excess noise
can also be appreciated from the analysis done in the Supplementary Section
IE6 of~\cite{CVMDIQKD} (in particular, see Fig.~5 there), where the security
thresholds are proven to be robust against much higher excess noise
($\varepsilon=$ $0.1$).

Then, Xu et al.\cite{Fehiu} completely missed an important advantage of
CV-MDI-QKD, which is the extremely good performance of CV\ detection performed
at the relay. This crucial feature relies on two basic facts:

\begin{description}
\item[(1)] The detection setup of CV-MDI-QKD is completely different from that
of one-way CV-QKD\ protocols.

\item[(2)] CV Bell detection is deterministic and highly efficient (also in
telecom setups).
\end{description}

\begin{figure*}[ptbh] \vspace{-3cm}
\par
\begin{center}
\includegraphics[width=0.99\textwidth] {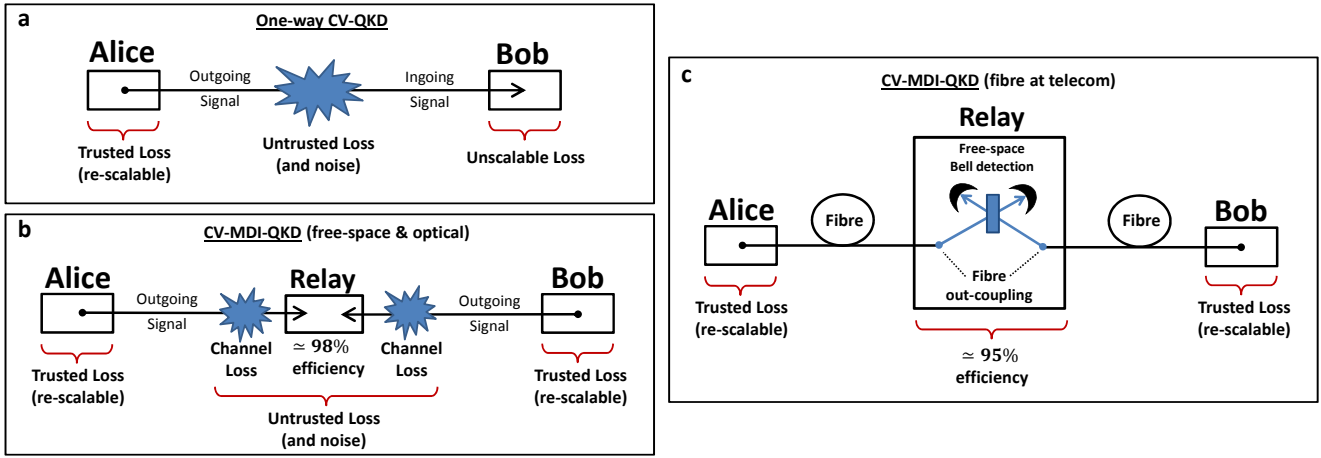}
\end{center}
\par
\vspace{-3.4cm}\caption{Experimental loss in CV-QKD configurations
(see text for more details). \textbf{a}, One-way CV-QKD protocol
with limited efficiency within Bob's station, due to unscalable
loss. \textbf{b}, Proof-of-principle demonstration of
CV-MDI-QKD~\cite{CVMDIQKD} in free-space at optical/infrared
regime ($1064$~nm) with $\simeq98\%$ efficiency for the CV Bell
detection. \textbf{c}, Proposal for a fibre-optic telecom
implementation of CV-MDI-QKD, where fibre-outcoupling and
free-space CV Bell detection achieve
$\simeq95\%$ efficiency.}%
\label{PMscheme}%
\end{figure*}

\noindent Let us explain these points in detail and the reader may also refer
to the panels in Fig.~\ref{PMscheme}.

In one-way (or point-to-point) CV-QKD protocols (see
Fig.~\ref{PMscheme}a), Alice prepares outgoing quantum states
while Bob detects states incoming from the channel. Because of
this configuration, only the loss within Alice's station can be
trusted and, therefore, neglected by re-scaling the signal level
at the output of Alice's box. The loss within Bob's station cannot
be re-scaled since it is added on top of channel loss and noise
(any re-scaling will also amplify the loss and noise of the
channel, without any signal-to-noise advantage). For this reason,
Bob's overall quantum efficiency is limited and this clearly
affects fibre-optic implementations at telecom wavelengths. For
instance, a quantum efficiency of $60\%$ was reported in
Ref.~\cite{Grangier2} due to optical manipulations and
photo-detection (however, note that this specific value of $60\%$
can be improved and, therefore, should not be considered as a
fundamental limit for the detection in one-way CV-QKD
protocols).

By contrast, in the setup of CV-MDI-QKD depicted in Fig.~\ref{PMscheme}b,
\textit{both} Alice and Bob prepare outgoing quantum states in their private
stations, where loss and noise are trusted. For this reason, both the signal
levels of Alice and Bob can be set at the output of their stations, so that
the effect of the internal losses can be completely neglected. Untrusted loss
and noise will only affect the channels and the detection at the relay. The
latter has extremely high efficiencies, indeed $\eta_{d}\simeq98\%$ in our
free-space experiment at optical wavelengths ($1064$ nm).

Here it is worth remarking that the CV version of\ Bell detection can be done
deterministically with simple linear optics and
photodetection~\cite{telereview}, contrarily to the $25\%$ probability success
affecting DV-MDI-QKD and the\ typical $50\%$ value bounding teleportation
experiments with photonic qubits~\cite{telereview,Bell1,Bell2,Bell3}. Most
importantly, the quantum efficiency of the homodyne detectors in CV Bell
detections is extremely high. Contrary to the incorrect claims of Xu et
al.~\cite{Fehiu}, homodyne detectors with high efficiencies have been seen in
both optical and telecom setups:

\begin{description}
\item[{[Optical Setups]}] These include the various free-space CV
teleportation experiments~\cite{telereview}, but also fibre-based experiments,
such as Ref.~\cite{Yone,Yoko}, where coupling efficiency to a fibre can be as
high as $98-99\%$ and the quantum efficiency of photo-detectors can be
$>99\%$, with an overall efficiency of about $97-98\%$~\cite{Akira}.

\item[{[Telecom Setups]}] For instance, see Ref.~\cite{Tobias2}, where the
performance of a balanced homodyne detection at $1550~$nm is $98\%$, by means
of InGaAs PIN diodes with an active area of $500~\mu$m in diameter and a
quantum efficiency of $\sim99\%$, together with a fringe visibility of
$99.5\%$ at the beamsplitter. See also Ref.~\cite{Tobias1}, where the overall
efficiency of the balanced homodyne detector at $1550~$nm is $95\%$. This is
performed in free space at the output of a $10$~m fibre, and the total $95\%$
efficiency includes both fibre in-coupling (the largest source of loss) and
outcoupling (estimated to be $0.1\%$)~\cite{Tobiasnote}.
\end{description}

A simple fibre-optic implementation of CV-MDI-QKD at telecom
wavelengths can be done as shown in Fig.~\ref{PMscheme}c. Here the
loss within Alice's and Bob's stations (e.g., associated with
fibre connections and modulators) can all be re-scaled as trusted
loss. At the relay, the CV Bell detection can be done in free
space. The efficiency associated with fibre out-coupling is close
to $100\%$ if the fibre facet is anti-reflection coated. The
efficient of the subsequent CV Bell detection is basically
determined by the homodyne detectors which may have $98\%$
efficiency at $1550~$nm~\cite{Tobias2}. For instance, the CV Bell
detection can be done in the simplified setup of
Ref.~\cite{CVMDIQKD} involving a balanced beam-splitter and two
photodectors. Considering $0.2$ dB insertion loss for the beam
splitter and $99\%$ efficiency for the diodes~\cite{Tobias2}, one
realizes a CV Bell detection with $\simeq95\%$ efficiency. This
means that the performance of our proof-of-principle experiment
can be achieved in a future fibre-optic telecom
implementation, contrary to the conjectures made by Xu et al.~\cite{Fehiu}%
\thinspace\ where the protocol was analysed by assuming too low efficiencies
for the detection at the relay (down to $85\%$).

Unfortunately, yet other claims made by Xu et al.~\cite{Fehiu} were wrong.
Contrarily to what they state:

\smallskip

\noindent\textbf{(i)~The asymptotic rate of~\cite{CVMDIQKD} is not an upper
bound but a lower bound with respect to all possible attacks}. In fact, the
experimental rate is computed from Alice and Bob's shared classical data
assuming the whole environment belongs to Eve. Then, the theoretical rate is
derived against optimal attacks. More precisely, this is minimized over all
two-mode Gaussian attacks in normal form after the application of the de
Finetti theorem and the extremality of Gaussian states~\cite{RMP}. This is in
contrast with the partial (and purely numerical) analyses in~\cite{Li,MaXC}
which only considered the simple case of two independent entangling cloner
attacks, as thoroughly discussed in Ref.~\cite{Carlo}. More details may be
found in~\cite{Notaprova}.

\smallskip

\noindent\textbf{(ii)~Finite-size effects against coherent
attacks~\cite{Jouguet,Elkouss,RMP} and composable security in the presence of
collective attacks~\cite{Leverrier} support our results}. It is true that our
theoretical rate is derived in the asymptotic limit of infinite signals but
there is no evidence that it would be sensibly affected by finite-size
analyses. In fact, the current finite-size analyses tend to the asymptotic
limit for blocks of $\gtrsim10^{8}$ data points (e.g., see Fig.~1\ of
\cite{Leverrier}), which is the size considered in
our\ experiment~\cite{CVMDIQKD}. This is also the size of data blocks
considered in the experiment of Ref.~\cite{Grangier2}, where the finite-size
key rate is shown to well-approximate the asymptotic regime, especially below
$30$~km, as is clear from Fig.~2 of \cite{Grangier2}. Note that
(point-to-point) coherent-state protocols are composable-secure against
general coherent attacks, but the present proof techniques are not sufficient
to prove fast convergence~\cite{Leverrier2,RevDIA}. The fact that a proof is
currently missing does not mean that such convergence cannot be shown with
another method, i.e., there is no fundamental reason to conjecture that the
composable-secure key rates of\ coherent-based CV-QKD\ protocols should not
fastly converge to the asymptotic values. Indeed this fast convergence has
been already proven in the most general case for squeezed-state
CV-QKD~\cite{RevDIA}. By contrast, we note that finite-size analysis seems to
be very demanding\ for DV-MDI-QKD, where blocks of $\gtrsim10^{10}$ data
points are needed for achieving reasonably non-zero rates with practical
detectors (see Fig.~4 of~\cite{CurtyFIN}).

\smallskip

\noindent\textbf{(iii)~The relay doesn't need to be in Alice's lab}. In fact,
all configurations (symmetric or asymmetric) show a non-trivial advantage with
respect to DV-MDI-QKD at metropolitan distances ($5-25$km).

\smallskip

\noindent\textbf{(iv)~The use of a single local oscillator is not a major
security flaw in CV-QKD and its removal is no longer an experimental
challenge}. The use of a single local oscillator is typical in all CV-QKD
implementations so far. Practical security against its potential manipulation
can be achieved if one implements an accurate real-time measurement of the
experimental shot noise~\cite{Loscill,RevDIA}. Furthermore, CV-QKD can also be
implemented using two independent local oscillators followed by classical
post-processing~\cite{Qi2,Soh,NoteREMAP}. Thus, the \textquotedblleft source
requirements\textquotedblright\ brought up by Xu et al.\ (Appendix~D
of~\cite{Fehiu}) are easily overcome. Regarding this issue of using a single
laser, it is important to remark that Xu et al.'s comment~\cite{Fehiu} is not
just against CV-MDI-QKD but the entire field of CV-QKD. According to Xu et
al.'s `criteria', all CV-QKD protocols implemented so far, including the first
ground-breaking table-top experiment of Ref.~\cite{Grangier} and the
long-distance experiment of Ref.~\cite{Grangier2} wouldn't be, as they put it,
\textquotedblleft properly designed QKD\ demonstrations\textquotedblright\ for
their use of a single local oscillator.

\smallskip

\noindent\textbf{(v)~Fast homodyne detectors exist plus CVs remain
superior even on slower clocks}. In the optical range, homodyne
detection can be done at GHz-bandwidths~\cite{HomGHZ}, with
$80$MHz detectors being available with efficiencies
$\gtrsim86\%$~\cite{BHD80} and, more recently, $100$MHz detectors
at $99\%$ efficiency~\cite{AkiraLAST}. This is a technology which
has large room of development at telecom wavelengths, where a
field implementation of CV-QKD has been already achieved at $1$MHz
clock~\cite{Grangier2}. This scenario has to be compared with the
$75$MHz clock rate implemented in DV-MDI-QKD~\cite{Tang}, while
$1$GHz has only been used for point-to-point BB84, with detector
efficiencies of about $20\%$~\cite{Luca} (now increasable up to
$55\%$~\cite{Shields1}). Here it is important to note that the
difference, in terms of bits per use, can be so large that CVs may
achieve higher rates than DVs while using much slower clocks. For
instance, in the case of point-to-point QKD, a $50$MHz-clock CV
protocol is already sufficient to outperform a $1$GHz-clock
protocol with DVs at metropolitan distances~\cite{Elkouss}. Since
there are about three orders-of-magnitude in the rate (bits per
use) between practical CV-MDI-QKD and practical DV-MDI-QKD, the CV
protocol could run on clocks which are three orders-of-magnitude
slower and still achieve the same performances of DVs. For this
reason, a fibre-based telecom implementation of CV-MDI-QKD at
$1~$MHz is hard to beat by any practical implementation of
DV-MDI-QKD (see below for a full clarification of the word
\textquotedblleft practical\textquotedblright\ in this
context).\begin{figure*}[ptbh] \vspace{-0.2cm}
\par
\begin{center}
\includegraphics[width=0.99\textwidth] {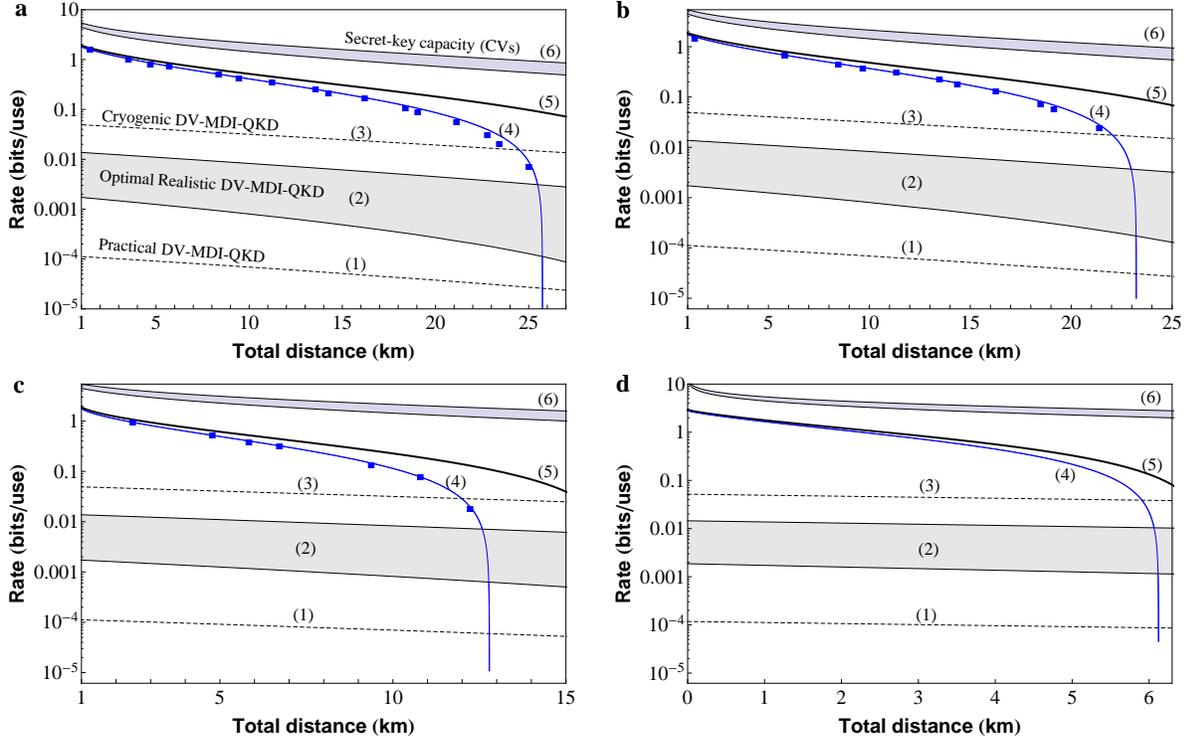}
\end{center}
\par
\vspace{-0.9cm}\caption{Comparison between the rates (bits per
relay use) of CV-MDI-QKD and DV-MDI-QKD versus total Alice-Bob
distance (in fibre-equivalent at 0.2dB/km) and assuming different
configurations for the relay in the various panels
\textbf{a}-\textbf{d}. \textbf{a}, relay in Alice's lab. From
bottom to top we consider: (1) Rate of practical DV-MDI-QKD\ with
standard semiconductor detectors ($\eta_{d}\simeq14.5\%$,
$e_{d}\simeq1.5\%$, $Y_{0}\simeq6\times10^{-6}$, $f_{e}=1.16$).
(2) Rate of the optimal realistic DV-MDI-QKD\ with the best
semiconductor detectors ($\eta_{d}\simeq55\%$,
$Y_{0}\simeq5\times10^{-4}$). Lower bound of the gray band refers
to typical error $e_{d}\simeq1.5\%$, while the upper bound refers
to $e_{d}\simeq0.1\%$ ($f_{e}=1.16$ in all cases) (3) Rate of
ideal DV-MDI-QKD with cryogenic
93\%-efficiency SNSPDs ($\eta_{d}\simeq93\%$, $e_{d}\simeq0.1\%$, $Y_{0}%
\simeq10^{-6}$, $f_{e}=1.16$). (4) Rate of practical CV-MDI-QKD
with room-temperature cheap components ($\eta_{d}\simeq98\%$,
$\varepsilon \simeq0.01$, $\varphi\simeq60$, $\xi=97\%$); squares
are experimental data. (5) Rate of CV-MDI-QKD with ideal
reconciliation efficiency (as before but $\xi=100\%$). (6)
Secret-key capacity of the total channel between Alice and Bob,
lowerbounded by~\cite{Pirandola2009,reverse} and upperbounded
by~\cite{TGW}. \textbf{b}, as in \textbf{a}\ but with a 100m-fibre
between Alice and the relay. \textbf{c}, as in \textbf{a}\ but
with 1km-fibre between Alice and the relay. \textbf{d}, symmetric
case where Alice and Bob are equidistant (in simulated fibre) from
the relay; no experimental data for this specific configuration.}
\label{figure}
\end{figure*}

\bigskip

\section*{Actual comparison between CVs and DVs}

Having refuted the various incorrect claims and conjectures made by Xu et al.
regarding CVs, we now perfom a detailed comparison between CVs and DVs,
considering a realistic analysis of DV-MDI-QKD. We can easily see that
practical parameters and cheap components for DVs are certainly \textbf{not}
those stated in Xu et al.'s comment~\cite{Fehiu} but those considered by four
of these same authors in an earlier paper on \textquotedblleft practical
aspects\textquotedblright\ of DV-MDI-QKD~\cite{Xu}. Practical DV-MDI-QKD
involves standard single-photon detectors with $\eta_{d}\simeq14.5\%$
efficiency and a $Y_{0}\simeq6\times10^{-6}$ dark count rate, together with
typical values of the intrinsic error rate $e_{d}\simeq1.5\%$ and error
correction efficiency parameter $f_{e}=1.16$~\cite{Brassard}. This is
state-of-the-art for room-temperature (20$%
{{}^\circ}%
$C) or thermoelectrically cooled (-50$%
{{}^\circ}%
$C) semiconductor (InGaAs) avalanche photodiodes (APD)
detectors~\cite{Shields2,Zbinden}, typically operating at $10-30\%$ efficiency
but with increasing dark counts (see Fig.~2 of~\cite{Shields2}).

Assuming these realistic and practical parameters, one can easily compute an
estimate of the rate for DVs by assuming infinite decoy states, neglecting
finite-size effects and other errors such as time-jitter and mode mismatch
(see Appendix for details). From Fig.~\ref{figure} we see how the rate of the
practical DV-MDI-QKD is below that of the practical CV-MDI-QKD by $3-4$
orders-of-magnitude over metropolitan distances, from $6$km in the symmetric
case up to $25$km in the most asymmetric configuration.

It is interesting to study the maximum rate that one can achieve
with DVs by using semiconductor detectors. Since we are interested
in short to medium distances and high rates, the choice is for
detectors with the highest efficiency (despite their high dark
counts). The best solution appears to be the very recent
InGaAs/InP APDs operated in optimized self-differencing mode at
room temperature by Toshiba~\cite{Shields1}. These achieve
$\eta_{d}\simeq55\%$ efficiency with background rates $Y_{0}\simeq
5\times10^{-4}$ at 20$%
{{}^\circ}%
$C. Fixing $f_{e}=1.16$ and considering $e_{d}$ in the range $0.1\%-1.5\%$ we
can estimate the optimal performance of realistic DV-MDI-QKD in the lower gray
band shown in Fig.~\ref{figure}. As we see from the figure, the optimal
DV-MDI-QKD with semiconductor detectors is very good but still between 2 and 3
orders of magnitude below the already experimentally achieved CV rate over
typical metropolitan ranges.

In their comment, Xu et al.~\cite{Fehiu} compared our practical CV
experiment with a purely-theoretical numerical extrapolation for
DVs, which is based on nearly-ideal parameters and devices. In
particular, they consider an extremely-low intrinsic error rate
$e_{d}\simeq0.1\%$~\cite{Tang} (which would be very demanding to
realise with real-time modulators in a scalable practical
network)\ and the most advanced superconducting nanowire
single-photon detectors (SNSPDs). These detectors operate below 2K
with $\eta_{d}\simeq93\%$ efficiency and a
$Y_{0}\simeq6\times10^{-6}$ background rate~\cite{Marsili}. It is
interesting to note that two of the authors of Xu et
al.~\cite{Fehiu} seem to be aware of the limitations of the
applicability of these SNSPDs to a
real practical scenario, as they openly admitted in a previous publication:%
\begin{align*}
&  \text{\textquotedblleft The main drawback of these novel SNSPDs, }\\
&  \text{however, is their operating temperature, }\\
&  \text{which is currently of the order of }0.1\text{~K.\textquotedblright}\\
&  \text{[Lo, Curty \& Tamaki, \textit{Nature Photon. }\textbf{8},
595-604
(2014)]}%
\end{align*}
It is evident that such an ideal and cryogenic version of DV-MDI-QKD is too
demanding for the realistic construction of a scalable network (more details
may be found in~\cite{Notareti}). Remarkably, even considering these
over-optimistic parameters, our practical CV protocol is still $\simeq1$
order-of-magnitude better than this ideal extrapolation with DVs in all the
configurations of Fig.~\ref{figure}.

Furthermore, one can easily verify that the DV rate collapses down by another
$1-2$ orders-of-magnitude by employing more standard SNSPDs with $\eta
_{d}\lesssim45\%$~\cite{Tang}. By contrast, slight improvements in the
reconciliation efficiency of classical protocols of error correction and
privacy amplifications may provide further non-trivial gains for CVs. As we
see in Fig.~\ref{figure}, the CV theoretical rates with ideal reconciliation
($\xi=100\%$) is very close to the secret-key capacity of the total Alice-Bob
channel~\cite{CVMDIQKD}, achievable by CV-QKD protocols.

That being said, our analysis here does not want to discourage the
implementation of DV-MDI-QKD at metropolitan distances. While Xu et al.'s
cryogenic version of DV-MDI-QKD does not seem to be suitable for building a
realistic and low-cost metropolitan network, an implementation of DV-MDI-QKD
with room-temperature (or slightly-cooled) semiconductor-based single-photon
detectors is not only realistic but appealing, and may be the basis of a
secure quantum network with intermediate rates. We therefore strongly
encourage serious and careful research in this direction. It would also be
interesting to explore potential hybrid DV-CV approaches, as it is now
happening for other quantum protocols (e.g., see quantum
teleportation~\cite{telereview}).

In conclusion, our take-home message is the following: DV-MDI-QKD is very good
for long distances~\cite{Tang,EXP1,EXP2}, but its rate struggles to be
increased at various distances, thus motivating the proposal of alternate
strategies by the community~\cite{ignore}. By contrast,
CV-MDI-QKD~\cite{CVMDIQKD,Carlo} struggles with long distances but can
potentially provide much higher rates at the metropolitan range ($5-25$km),
extending what happens for point-to-point CV-QKD~\cite{Elkouss,ScaraniRMP}.
This statement must not come as a surprise since CV systems are fragile to
loss (limiting distance) but they can encode a lot of information thanks to
their theoretically infinite-dimensional Hilbert space~\cite{RMP}.
Furthermore, besides the use of cheap and practical devices, CVs can easily go broadband.

\bigskip

\textbf{Acknowledgments.} Authors thank feedback from J. H. Shapiro, A.
Furusawa, Z. Zhang, H. Zbinden, F. Grosshans, A. Leverrier, and E. Diamanti.
S.P. also would like to thank R. Filip, V. Usenko, N. L\"{u}tkenhaus, Q.
Zhang, and V. Scarani for comments/discussions on CV-MDI-QKD and CV-QKD. This
work has been supported by the EPSRC `UK Quantum Communications HUB'
(EP/M013472/1) and `qDATA' (EP/L011298/1), the Leverhulme Trust, the H. C.
{\O }rsted postdoc programme, and the Danish Agency for Science, Technology
and Innovation (Sapere Aude project).

\appendix

\section{Secret-key rates}

\subsection{Secret-key rate of DV-MDI-QKD}

An estimate for the secret key rate of polarisation-encoding DV-MDI-QKD can
easily be computed by assuming infinite decoy states and signals (i.e.,
ignoring finite-size effects and composability) and neglecting a series of
technical errors, such as time-jitter, spectral mismatch, pulse-shape mismatch
and fluctuations of the source intensities. This is effectively an upper bound
to the actual performance of DV-MDI-QKD and corresponds to the quantity
computed by \cite{Fehiu}, following~\cite{Lo,Xu}. This (over-estimated)
secret-key rate is given by%
\[
R_{\text{DV-MDI}}=P_{Z}^{11}Y_{Z}^{11}\left[  1-H_{2}(e_{X}^{11})\right]
-Q_{Z}~f_{e}~H_{2}(E_{Z}),
\]
where $P_{Z}^{11}=\mu_{A}\mu_{B}\exp[-(\mu_{A}+\mu_{B})]$ is the probability
that Alice and Bob emit a single photon, with $\mu_{A}$ ($\mu_{B}$) being the
intensity of Alice's (Bob's) signal; $Y_{Z}^{11}$ is the yield in the
$Z$-basis and $e_{X}^{11}$ is the error rate in the $X$-basis, assuming that
the parties send single-photon states; $Q_{Z}$ and $E_{Z}$ are, respectively,
the gain and the quantum bit error rate (QBER) in the $Z$ basis; $f_{e}$ is
the error correction inefficiency, and $H_{2}(x)=-x\log_{2}x-(1-x)\log
_{2}(1-x)$ is the Shannon entropy.

Parameters $Y_{Z}^{11}$,\ $e_{X}^{11}$, $Q_{Z}$ and $E_{Z}$, can be simulated
assuming a series of conditions, which further qualify the result to be an
upper bound of the actual rate. These include the modeling of the intrinsic
error $e_{d}$ by means of two unitaries at the input of the beam splitter at
the relay station~\cite{Xu}. Considering single-photon detectors with
efficiency $\eta_{d}$ and dark count rate $Y_{0}$, one may
write~\cite{Fehiu,Xu}
\begin{align*}
Y_{Z}^{11}  &  =(1-Y_{0})^{2}\left[  4Y_{0}^{2}(1-\eta_{A}\eta_{d})(1-\eta
_{B}\eta_{d})\right. \\
&  \left.  +2Y_{0}\left(  \eta_{A}\eta_{d}+\eta_{B}\eta_{d}-\frac{3\eta
_{A}\eta_{B}\eta_{d}^{2}}{2}\right)  +\frac{\eta_{A}\eta_{B}\eta_{d}^{2}}%
{2}\right]  ,\\
e_{X}^{11}  &  =\frac{1}{2}-\frac{(1-Y_{0})^{2}(1-e_{d})^{2}\eta_{A}\eta
_{B}\eta_{d}^{2}}{4Y_{X}^{11}},
\end{align*}
where $Y_{X}^{11}=Y_{Z}^{11}$, and $\eta_{A}$ ($\eta_{B}$) is the
transmissivity of Alice's (Bob's) link with the relay. Then, one may write%
\[
Q_{Z}=\frac{\Omega_{1}+\Omega_{2}}{2},~~~E_{Z}=\frac{\Omega_{1}}{\Omega
_{1}+\Omega_{2}},
\]
where $\Omega_{1}$ and $\Omega_{2}$ are given in Eqs.~(A4) and~(A5)
of~\cite{Fehiu}.

Assuming various choices for the basic parameters $\eta_{d}$, $Y_{0}$ and
$e_{d}$, (with $f_{e}=1.16$~\cite{Brassard}), we maximize $R_{\text{DV-MDI}}$
over the intensities $\mu_{A}$ and $\mu_{B}$, deriving the simulated
theoretical rates shown in Fig.~\ref{figure}\ of the main text.

\subsection{Secret-key rate of CV-MDI-QKD}

The security proof of CV-MDI-QKD needs a dedicated discussion for
infinite-dimensional systems which involves various elements, including the
\textquotedblleft de Finettization\textquotedblright\ of the classical data
and the extremality of Gaussian states (see~\cite{CVMDIQKD} for details). From
Alice's amplitudes $\alpha$, Bob's amplitudes $\beta$ and the relay outcomes
$\gamma$, we can derive a joint classical probability $p(\alpha,\beta,\gamma)$
which identifies a conditional \textquotedblleft post-relay\textquotedblright%
\ state $\rho_{ab|\gamma}$ shared by Alice and Bob in the entanglement-based
representation of the protocol. Such a state is then purified into an
environment $E$ which is assumed to be fully controlled by Eve. From
$\rho_{ab|\gamma}$, we can derive both Alice and Bob's mutual information
$I_{AB}$, and Eve's Holevo information $\chi_{E}$ (e.g., on Alice's variable
$\alpha$). As a result, we may write the rate as
\[
R_{\text{CV-MDI}}=\xi I_{AB}-\chi_{E},
\]
where $\xi$ is the reconciliation efficiency of the classical codes for error
correction and privacy amplification. This is the general method adopted to
compute our experimental rate.

In order to derive the theoretical rate, we have to model the most general
Gaussian attack against the two channels, which is compatible with observed
channel transmissivities, $\eta_{A}$ and $\eta_{B}$. This is done by combining
the two lossy channels in the most general way. Since these canonical
forms~\cite{RMP} always admit a local (non-Stinespring) dilation with a beam
splitter and a thermal mode~\cite{RMP}, the two-mode Gaussian attack can be
represented by two beam splitters subject to a generally-correlated two-mode
Gaussian state for the environment.

By means of local displacements and symplectic transformations, this state can
be reduced to a zero-mean Gaussian state whose covariance matrix is in normal
form $\mathbf{V}(\omega_{A},\omega_{B},g,g^{\prime})$, where $\omega_{A}$
($\omega_{B}$) is the thermal noise affecting Alice's (Bob's) link, and
parameters $g$ and $g^{\prime}$ describe the correlations between Eve's modes.
In these conditions, Alice and Bob's mutual information can be written as
$I_{AB}=\log_{2}[(\varphi+1)\chi^{-1}]$, where $\varphi$ is the modulation of
the coherent states and $\chi$ is the equivalent noise, decomposable as
$\chi=\chi_{\text{loss}}+\varepsilon$, with $\chi_{\text{loss}}$ being the
pure-loss noise and $\varepsilon(\eta_{A},\eta_{B},\omega_{A},\omega
_{B},g,g^{\prime})$ the `excess noise'. For any fixed value of the
transmissivities and excess noise, we then optimize Eve's Holevo information
over the remaining degrees of freedom, i.e., $\omega_{A}$, $\omega_{B}$, $g$
and $g^{\prime}$. Thus, we compute a lower bound of the rate, denoted by
$R_{\varphi,\xi}(\eta_{A},\eta_{B},\varepsilon)$~\cite{CVMDIQKD}. This
quantity is asymmetric and decreases more rapidly in $\eta_{A}$ than in
$\eta_{B}$. For this reason, we obtain a further lower bound if we replace
$\eta_{A}\rightarrow\eta_{A}\eta_{d}$, where $\eta_{d}$ takes into account of
the overall efficiency of the CV Bell detection at the relay. This quantity is
here used to compute the theoretical rates shown in the main text.

Note that a simple analytical formula can be written in the case of ideal
reconciliation $\xi=1$ and large modulation $\varphi\gg1$. In this case, we
have~\cite{CVMDIQKD}
\begin{align*}
R(\eta_{A},\eta_{B},\varepsilon)  &  =\log_{2}\left[  \tfrac{2(\eta_{A}%
+\eta_{B})}{e|\eta_{A}-\eta_{B}|\chi}\right]  +\\
&  h\left[  \tfrac{\eta_{A}\chi}{\eta_{A}+\eta_{B}}-1\right]  -h\left[
\tfrac{\eta_{A}\eta_{B}\chi-(\eta_{A}+\eta_{B})^{2}}{|\eta_{A}-\eta_{B}%
|(\eta_{A}+\eta_{B})}\right]  ,
\end{align*}
where $\chi=2(\eta_{A}+\eta_{B})/\eta_{A}\eta_{B}+\varepsilon$ and%
\[
h(x):=\frac{x+1}{2}\log_{2}\frac{x+1}{2}-\frac{x-1}{2}\log_{2}\frac{x-1}{2}.
\]
In the symmetric case $\eta_{A}=\eta_{B}:=\eta$, this rate becomes%
\[
R(\chi)=h\left(  \frac{\chi}{2}-1\right)  +\log_{2}\left[  \frac{16}{e^{2}%
\chi(\chi-4)}\right]  .
\]

\subsection{Secret-key capacity}

The total transmissivity of an equivalent point-to-point lossy channel between
Alice and Bob is equal to $\eta_{\text{tot}}=\eta_{A}\eta_{B}$. The maximum
secret-key rate which is achievable by a CV-QKD protocol is the secret-key
capacity $K$ of the channel, satisfying~\cite{Pirandola2009,reverse,TGW}
\[
\log_{2}\left(  \frac{1}{1-\eta_{\text{tot}}}\right)  \leq K\leq\log
_{2}\left(  \frac{1+\eta_{\text{tot}}}{1-\eta_{\text{tot}}}\right)  .
\]

\section{Reply to the \textquotedblleft Appendix E: Addendum\textquotedblright%
\ of Xu et al.}

Here we provide an additional reply to Xu et al.\ who added an appendix to
their comment (see the second version available at
http://arxiv.org/abs/1506.04819v2). Most of the new comments are just
repetitions of the old arguments and have been already replied to in our text
above. However, we provide here, and again, a point-to-point rebuttal of these
issues, also claryfing other new incorrect statements made by these authors.
We apologize with the reader for the repetitions triggered by this process. It
is clear that Xu et al. are trying to make all their possible arguments
against CV-QKD once they realised they were not able to approach the rate of CV-MDI-QKD.

\begin{description}
\item[$\square1$] Xu et al.\ claim: \textquotedblleft Pirandola et al.\ agree
with our main point.\textquotedblright

\item[$\blacksquare1$] Our reply: \textbf{No, we do NOT agree with their main
point, and we are NOT contradicting our Nature Photonics paper~\cite{CVMDIQKD}%
}. Xu et al. are clearly mis-quoting us. As already explained before, the rate
of CV-MDI-QKD with practical devices and parameters is at least three
orders-of-magnitude higher than that of DV-MDI-QKD with corresponding
practical devices and parameters~\cite{Xu}. The CV-rate also keeps an
advantage of between two and three orders-of-magnitude with respect to the
optimal realistic implementation of DV-MDI-QKD with the best available
semiconductor single-photon detectors. This advantage is quantified in terms
of bits per relay use and evaluated for various values (in dBs) of the channel
loss. That being said, the theoretical possibility of using DV-MDI-QKD with
cryogenic devices (SNSPDs) currently seems: Unnecessarily complex and very
expensive; hard to miniaturise; not extendable to more complex networks (e.g.,
where each node may act as a user or as a relay); not extendable to mobile
devices (which can be hot-spots as well); very fragile with respect to the
performances of the SNSPDs, easily losing one-two orders of magnitude from
$>90\%$ to $<40\%$ efficiencies. Most importantly, better performances could
be achieved with practical implementations of CV-MDI-QKD with cheap and
room-temperature devices. Unfortunately, Xu et al. keeps comparing our
practical experimental data with the theoretical simulation of a potential
cryogenic implementation.

\item[$\square2$] Xu et al.\ claim: \textquotedblleft Experimental results of
CV-MDI-QKD done in free-space, at a non-telecom wavelength, and using
nontelecom detectors cannot and should not be used as a demonstration of
telecom CV-MDI-QKD performance.\textquotedblright\ Plus other redundant
statements about the use of a single laser etc.

\item[$\blacksquare2$] Our reply: Despite the fact that ours is a
proof-of-principle experiment in free space at $1064$~nm, it gives
a clear indication of the potential performances of the CV-MDI-QKD
protocol in a fibre-based telecom-wavelength implementation, where
highly efficient homodyne detectors are available at $1550$~nm and
fibre-couplings are not an issue. This is already fully explained
in the main text. In particular, see our points (1) and (2) in the
main text, their explanations and Fig.~\ref{PMscheme}. The use of
a single laser is not a major issue as already extensively
explained in point (iv) of the main text. \textbf{Therefore, our
experimental results can be used to make quantitative statements
about the future performance of a field implementation of
CV-MDI-QKD.}

\item[$\square3$] Xu et al.\ claim: \textquotedblleft CV-MDI-QKD could have an
advantage over DV-MDI-QKD only under rather restrictive conditions. We do not
deny that CV-MDI-QKD might have an advantage over DV-MDI-QKD, but only in a
rather restrictive parameter space where a combination of
assumptions/conditions are simultaneously satisfied, namely, (a) asymptotic
key rate for an infinitely long key, (b) high-efficiency (well above 85\%)
homodyne detectors, (c) highly asymmetric configuration where the relay is
close to one of the two users, Alice or Bob, (d) low loss (i.e., short
distance).\textquotedblright

\item[$\blacksquare3$] Our reply: This is another claim by these authors which
is very easy to disprove. The various points have been already replied to
before but, in any case, we can again stress the reasons here.

\item \qquad(a)~As already explained in point (ii) of the main text,
finite-size effects support our experimental results in the sense that these
analyses tend to the asymptotic limit for blocks of $\gtrsim10^{8}$ data
points, which is the size considered in our\ experiment~\cite{CVMDIQKD}.
Despite the fact that the theoretical rate is asymptotic, there is no reason
to believe that it would be sensibly affected by finite-size analyses. Here
the main argument of Xu et al. is that composability security for
coherent-state protocols against coherent attack has not yet proven to
converge quickly to the asymptotic analysis. This is mainly a problem of
finding the correct proof technique, it cannot be conjectured as a fundamental
problem with CVs. As a matter of fact, this problem does not even exist for
squeezed-state protocols~\cite{RevDIA}.\ By contrast, it is known that larger
data blocks ($\gtrsim10^{10}$ points) are needed for DV-MDI-QKD with practical
room-temperature detectors.


\item \qquad(b)~Homodyne detectors have high efficiencies (well above 85\%)
both at optical and telecom wavelengths, free-space or coupled to
fibre. This is typical and already explained above. For instance
see the bullet points on page 2 of the main text. The value of
$85\%$ of Xu et al. is unreasonably too low.

\item \qquad(c)~As already stated in point (iii)\ of the main text and
completely clear from Fig.~\ref{figure}, the CV protocol can be run both in
symmetric and asymmetric configurations with superior performances.

\item \qquad(d)~Typical distances are within the metropolitan range. For
instance, see again Fig.~\ref{figure} where they range between $6$ and $25$ km.

\item[$\square4$] Xu et al.\ claim: \textquotedblleft For CV-QKD with coherent
states, practical composable secure key rates against the most general type of
attacks have yet to be shown [...] In this respect, Pirandola et al. [23]
confuse the restricted class of collective attacks with the most general class
of coherent attacks [...] current security proofs against the most general
type of coherent attacks for CV-QKD with coherent states deliver basically
zero key rate for any reasonable amount of signals\textquotedblright

\item[$\blacksquare4$] Our reply: This has been already replied to above.
\textbf{Not having a proof of fast convergence is one thing, proving that
there is no fast convergence is another.} Xu et al. are either confused by
this point, or they are trying to\ make a biased use of this argument.
\textbf{According to Xu et al, all the coherent-state protocols implemented so
far, including the ground-breaking results achieved in point-to-point
CV-QKD~\cite{Grangier,Grangier2}, would then run the risk of delivering zero
key rate in the composability security framework.}\ This implication seems to
be rather absurd, also considering the fact that, for squeezed-state
protocols, fast convergence has already been proven. And, yes...luckily we
know the difference between collective and coherent attacks (among other
things, some of our co-authors have defined the most general form of
collective Gaussian attack against CV-QKD protocols~\cite{CollATT}).

\item[$\square5$] Xu et al.\ claim: \textquotedblleft The critical dependence
of CV-MDI-QKD key rates on homodyne detection efficiency is down played by
Pirandola et al.\textquotedblright

\item[$\blacksquare5$] Our reply: \textbf{No}. This is just a repetition,
already replied to in (b) of point $\blacksquare3$ above.

\item[$\square6$] Xu et al.\ claim: \textquotedblleft Pirandola et al.'s
free-space experiment is not a properly designed QKD demonstration, thus the
results shown in Fig. 1 are all purely-numerical\textquotedblright. Their
reasons would be (a)~\textquotedblleft The use of a single
laser\textquotedblright; (b)~\textquotedblleft Table-top experiment at a
wavelength outside the telecom band\textquotedblright\ and
(c)~\textquotedblleft Incorrect noise model\textquotedblright\ (i.e., low
excess noise).

\item[$\blacksquare6$] Our reply: \textbf{No}. This is another repetition. Our
proof-of-principle experiment is a properly-designed\ free-space QKD
experiment, which provides clear indications of the full potentialities of the
protocol in a future field implementation with fibre connections at telecom
wavelengths, as already discussed in point $\blacksquare2$ above. In particular:

\item \qquad(a)~About the single laser: As already explained before, the
malicious manipulation of the local oscillator can be overcome with practical
experimental techniques~\cite{Loscill,RevDIA}. Note that all CV-QKD protocols
have been implemented with a common local oscillator, as is our CV-MDI-QKD
protocol. \textbf{Therefore, Xu et al. would automatically claim that all
CV-QKD protocols would not be \textquotedblleft properly designed QKD
demonstrations\textquotedblright.}

\item \qquad(b)~About the table-top experiment:\ Again, this is already
replied to in point $\blacksquare2$ above.

\item \qquad(c)~About the noise model: As already explained in the main text,
our experimental excess noise $\varepsilon\simeq0.01$ is not low but typical,
fully comparable with the values reported in the fibre-optic experiment
of~\cite{Grangier2}, where the experimental excess noise was estimated to be
$\varepsilon\simeq0.001$ at $10^{8\text{ }}$data points and $\varepsilon
\simeq0.008$ at $10^{6\text{ }}$data points, for Bob's detection at $53$~km.
At lower distances $\lesssim25$~km (as is in our case), the excess noise can
actually go below these values, which means that our experimental value
$\varepsilon\simeq0.01$ can even be considered relatively high. The robustness
of CV-MDI-QKD against excess noise can also be appreciated from the analysis
done in the Supplementary Section IE6 of~\cite{CVMDIQKD} (in particular, see
Fig.~5 there), where the security thresholds are proven to be robust against
much higher excess noise ($\varepsilon=$ $0.1$). Even assuming the numbers
given by Xu et al. ($0.015$), this is completely compatible with our
experimental value.
\end{description}

\bigskip

\begin{description}
\item[$\square7$] Xu et al.\ claim: \textquotedblleft CV-MDI-QKD could be
suitable only for particular network architectures\textquotedblright%
\ repeating some arguments on the sub-optimal performance of the symmetric configuration.

\item[$\blacksquare7$] Our reply: Xu et al. make several errors here:

\item \qquad(a) A real network is very unlikely to be perfectly symmetric, but
most of the nodes will be in an asymmetric configuration.

\item \qquad(b) The model of star network they consider is very restrictive.
This model is necessary if the relay needs to be implemented with very
expensive cryogenic detectors (as those suggested by these authors for
DV-MDI-QKD). By contrast, CV-MDI-QKD would allow one to use cheap detectors
for the relay, which means that every node in the network can play a role as a
user or as a relay, depending on the situation. In this more flexible
scenario, Alice just needs to connect to the closest (untrusted) node
available to secretly communicate with a remote (authenticated)\ Bob.

\item \qquad(c) Finally, even assuming the symmetric configuration, Alice and
Bob can be separated by a $6$ km-long fibre, which is not exactly a short
distance at the metropolitan scale.
\end{description}

At the end of their \textquotedblleft Addendum\textquotedblright\ Xu et
al.\ make some \textquotedblleft Final remarks\textquotedblright. Some of
these are completely secondary issues, others are just repetitions of previous
points which have been already addressed or refuted, especially the reiterated
points on the finite-size effects, the use of a single laser, and the homodyne
detectors. We note that Xu et al. misunderstood in which sense our theoretical
rate is \textquotedblleft optimal\textquotedblright, thinking that it would be
a sort of upper bound. The optimality is clearly intended as a minimization
over all possible attacks which therefore qualifies the key rate as a lower
bound -- see point (i) of the main text for full details. Thus, what Xu et al.
say (\textquotedblleft We admit that it was unclear to us whether or not the
realistic Gaussian attack considered in [1] to derive the theoretical rate was
indeed optimal, or if it provided an upper bound on the theoretical rate. So,
in our simulations we opted for the most favourable case for CV-MDI-QKD by
considering that the attack is optimal....We are glad to see that Pirandola et
al. [23] confirm this.\textquotedblright) is not confirmed by us and it is
clearly wrong.


\begin{thebibliography}{99}
\bibitem {Note}Note that Ref.~\cite{SidePRL}, which is the Letter preeceding
Ref.~\cite{Lo} in the same issue of PRL, independently introduced the idea of
end-to-end QKD with an untrusted relay, i.e., the scheme of
measurement-device-independence. Ref.~\cite{SidePRL} considered more general
DV systems (not just qubits) and more general quantum measurements performed
by the untrusted relay (not only Bell detections). Security was studied in the
entanglement-based representation.
\bibitem {Tobiasnote}Note that both the experiments in Refs.~\cite{Tobias2}
and~\cite{Tobias1} involved one of our co-authors (T. Gehring, formerly known
as T. Eberle).
\bibitem {Notaprova}Security can be studied in the entanglement-based
representation. Assuming this representation, the purification of Alice and
Bob's shared state into an environment which is fully-controlled by Eve is
only the first step of a theoretical analysis which aims at deriving an
analytical formula for the secret-key rate. In fact, it is generally
impossible to derive a simple closed formula in terms of all the parameters
describing Alice and Bob's shared state (in particular, their first- and
second-order moments in the Gaussian case). The next step is therefore to find
a lower bound to the key rate by minimizing over all the degrees of freedom
associated with Eve's dilation once a small set of accessible parameters has
been fixed in Alice and Bob's shared state (by ``accessible'' we mean always
computable from Alice and Bob's shared data). In CV-MDI-QKD, this small set is
given by the transmissivities of the two links and the overall excess noise
(easily computable from the shared data). By fixing these parameters, one has
to optimize Eve's attack over all the remaining degrees of freedom, including
the parameters describing the potential correlations between Eve's ancillary
modes. Thus, it is clear that just assuming the simple Markovian case of two
entangling cloners (i.e., absence of correlations) is a great limitation for
Eve and such strong hypothesis cannot provide an unconditionally-secure key rate.
\end{thebibliography}
\end{document}